# Tangled nonlinear driven chain reactions of all optical singularities

V.I. Vasil'ev, M.S. Soskin

Institute of Physics NAS of Ukraine, Prospect Nauky 46, Kyiv, Ukraine 680028

## ABSTRACT

Dynamics of polarization optical singularities chain reactions in generic elliptically polarized speckle fields created in photorefractive crystal LiNbO3 was investigated in details Induced speckle field develops in the tens of minutes scale due to photorefractive 'optical damage effect' induced by incident beam of He-Ne laser. It was shown that polarization singularities develop through topological chain reactions of developing speckle fields driven by photorefractive nonlinearities induced by incident laser beam. All optical singularities (C points, optical vortices, optical diabolos,) are defined by instantaneous topological structure of the output wavefront and are tangled by singular optics lows. Therefore, they have develop in tangled way by six topological chain reactions driven by nonlinear processes in used nonlinear medium (photorefractive LiNbO3:Fe in our case): C-points and optical diabolos for right (left) polarized components domains with orthogonally left (right) polarized optical vortices underlying them. All elements of chain reactions consist from loop and chain links when nucleated singularities annihilated directly or with alien singularities in 1:9 ratio. The topological reason of statistics was established by low probability of far enough separation of born singularities pair from existing neighbor singularities during loop trajectories. Topology of developing speckle field was measured and analyzed by dynamic stokes polarimetry with few seconds' resolution. The hierarchy of singularities govern scenario of tangled chain reactions was defined. The useful space-time data about peculiarities of optical damage evolution were obtained from existence and parameters of 'islands of stability' in developing speckle fields.

Keywords: singular optics, dynamics of singularities, optical vortex, polarization singularity, optical diabolos, local topological transition, chain topological reaction, loop topological reaction.

## 1. INTRODUCTION

Singularities are essential feature in many areas of science from Universe (strings), solids (crystals screw and edge dislocations), wave processes of any nature. Singular optics [1, 2] has shown that namely optical singularities define main features of optical wavefronts. New step of singular optics became nonlinear singular optics [3] especially realized in nonlinear media with singularities (optical solitons [4], especially in optical vortex solitons [5]). This paper prolongs started recently *dynamic* singular optics of generic developing singular optical fields created and driven by nonlinear processes in nonlinear media [6 – 12]. Chain and loop trajectories were found. Morphology of local reactions during singularities pair nucleations and annihilations were studied in details and their regularities were established. This paper is devoted to establishment of general regularities of tangled nonlinear driven chain reactions when all optical singularities are involved. Main emphases was made on the role of optical diabolos which arrange major and minor axes of polarization ellipses surrounding circularly polarized C points in two-cone structure [7-9, 13]. Optical diabolos arrangement and transformations give the most useful information about topology of generic developing light fields. Nucleation (annihilation) of optical diabolos and direct-feedback transformations of hyperbolics and elliptics changes instantaneously shape of correspondent speckle wavefront. This new effect was fixed firstly and investigated. The last task of our investigations was attempts extract useful information about dynamics of optical damage in used photorefractive crystal from obtained optical experimental data.

Each point on arbitrary wavefront is characterized by the Poincaré index $I_P$ equal phase change in $2\pi$ unites during circumference around it. Poincaré index equals +1 and −1 for extreme and saddles accordingly (Fig.1). $I_P = 0$ for all other points on the slopes of topological landscape with arbitrary complexity.

Usage and measurement of optical diabolos [2] (Fig.2) appear very helpful, because they give comprehensive information about topology of wavefront with arbitrary complexity and its transformations because shape and location of hyperbolics and elliptics on a speckle differ crucially [6-13]. As always, total Poincaré index is conserved during all transformations including singularities nucleation and annihilation by pairs [2]. Namely diabolos give needed information about evolution of wavefront topology in each moment of chain reactions because only they characterize the shape in each point of wave front. For hyperbolics (Fig.2*a*) phase changes in opposite direction during full circumference along its height contour due to knee on it located in C point and $I_P = 0$. Height contours near elliptics are smooth in any point and $I_P = 1$. Both forms of diabolo are generic ones, but elliptic is more structurally stable as usual extreme against hyperbolic which can easily glide on the speckles slops. Its position is not stable and flexible.

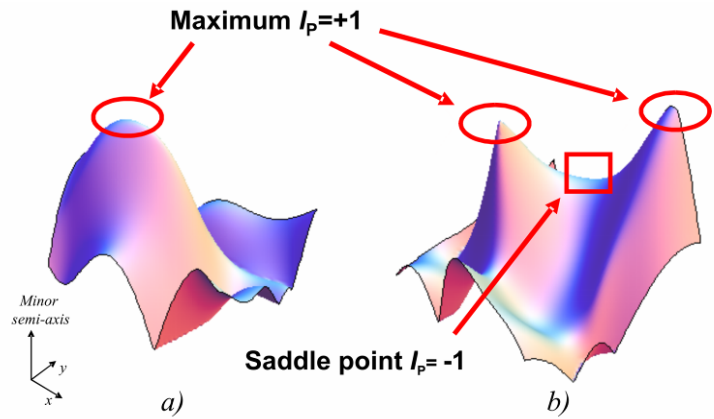

Figure 1. Critical points in arbitrary wavefront: a) local maximum with $I_P = +1$, b) saddle with $I_P = -1$.

From other site, pair of hyperbolics with total zero Poincaré index can be born in each point on wavefront slopes without violation of conservation low of total Poincaré index during any transformations. Due to this, they can move freely through the speckle field following its changes [9] like free carriers in superconductors. As you will see just now, this difference is crucial for topology of chain reactions. Nucleation of elliptics pair with total $I_P = +2$ which replace maximum of a speckle with $I_P = +1$, can be realized only by nongeneric splitting of speckle top on two tops with a saddle between them ($I_P = -1$) [14].

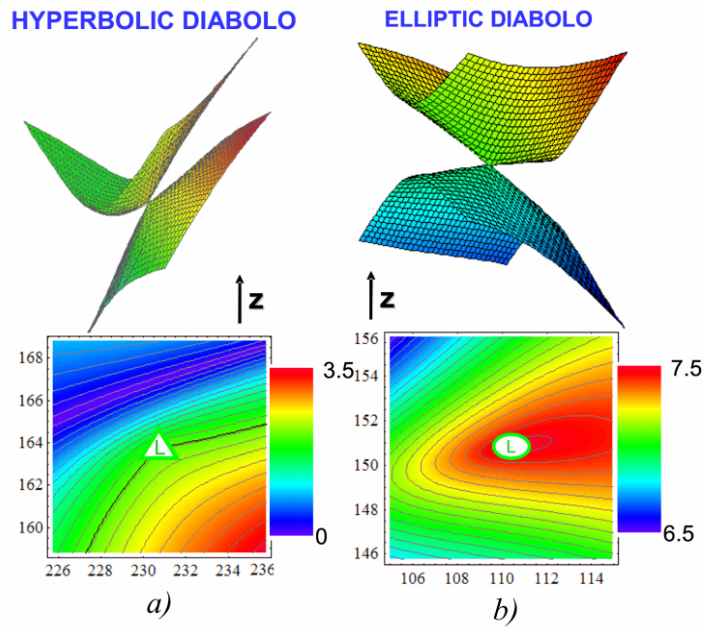

Figure 2. Hyperbolic (a) and elliptic (b) optical diabolos with two cones of polarization ellipses axes up and below C point. Upper (lower) cone corresponds to distribution of major (minor) axis of polarization ellipses in each point of wavefront. Poincaré index IP of hyperbolic (elliptic) diabolo equals 0 (+1). It is seen part of thick level height contour of shown hyperbolic which embraces top of the local speckle. Elliptic is always located on the top of a speckle. Triangle (ellipse) mark hyperbolic (elliptic) diabolos. Initial letter L marks lemon morphological form of long axes arrangement in polarization ellipses surrounding C point.

It's clear that this three-body reaction is nongeneric one. Indeed, we haven't fixed it in between thousands of measured events of diabolos nucleation. The actual topological properties of optical diabolos are next [13]: (*i*) the cones of optical diabolo are generically anisotropic, but the diabolo itself is always centrosymmetric; (*ii*) the only source of elliptics cones in optical fields is transformation of nearest hyperbolics; (*iii*) hyperbolic cones can, and do, migrate throughout the changeable wavefront. All local scenarios of diabolos birth and transformation predicted in [13] were confirmed experimentally [12].

Next general topological lows govern optical diabolos nucleation, transformation and annihilation in developing light fields:

- The generic nucleation (annihilation) of hyperbolics pair H (S) – H (M) is possible in each point on speckles slope because total Poincaré index of HH pair equals $I_P = 0$;

- Hyperbolic climbed on the top of a speckle transforms generically to elliptic (**H → E**) without change of total $I_P$ index because top of a speckle possesses $I_P = 0$ and its transformation to elliptic don't change the total $I_P$;

- Opposite transformation E → H is realized during elliptic descend from top of a speckle without change of total $I_P$ index also because $I_P = +1$ are valid both for elliptic and speckle top.

As will be shown below, experimental results were in full agreement with these lows.

## 2. EXPERIMENTAL TECHNIQUE

Subject of our investigations was dynamic topology of speckle field created by scattering of propagating laser beam and changeable generically due to 'optical damage' effect in photorefractive crystals LiNbO$_3$: Fe (Fig.3*a*) [15].

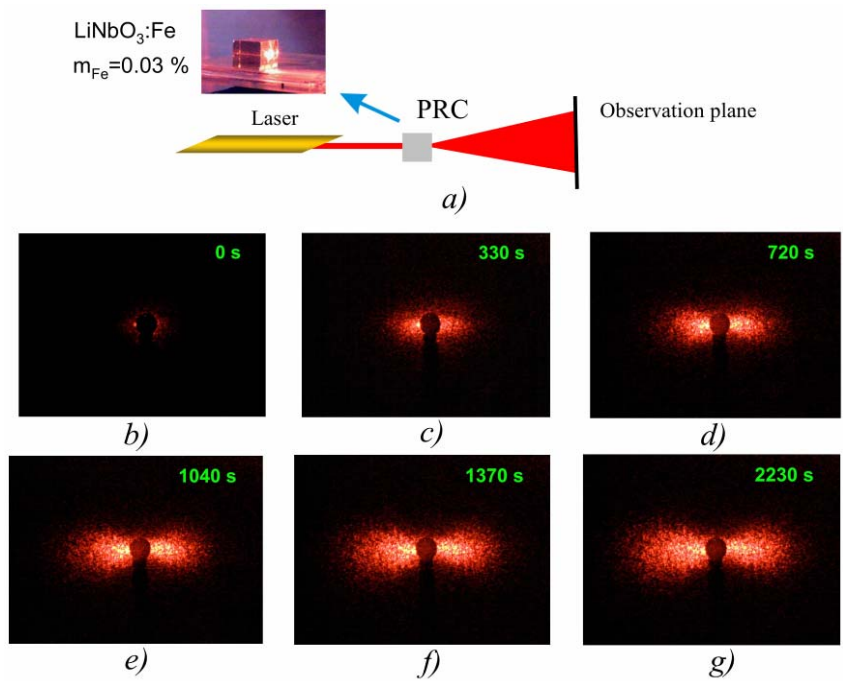

Figure 3. a) - arrangement for realization of "optical damage" effect, b) - starting structure of screened output laser beam with extra low scattering on initial refractive index heterogeneities, c) - g) - time development of speckled scattering full with optical singularities induced by propagating laser beam due to the 'optical damage' effect in LiNbO$_3$.

Optical damage develops slowly enough and saturates after nearly one hour of irradiation (Fig.3*b-g*). As we shown before, appeared speckle field contains multitude of optical singularities and develops generically together with induced speckle field by movement, nucleation and annihilation phenomena [10 – 12]. Such developing speckle field appeared unique and single up to moment polygon for inspection of theoretical predictions (for example [13, 15]) and establishment of genuine regularities of optical singularities dynamics.

Full experimental setup is presented in Fig.4.

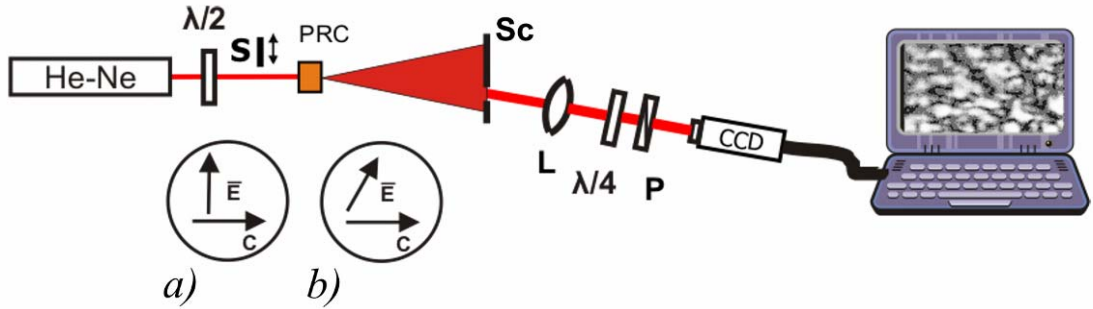

Figure4: Scheme of experimental setup.

Creation of time-dependent singular vector (elliptically polarized) speckle fields was realized in the one-beam setup by elaborated two-step procedure. The scalar gratings are written in crystal on the first step. Radiation emission with a wavelength of 0.63 nm generated by a helium-neon laser He-Ne passed through $LiNbO_3$:Fe crystal PRC doped with iron to 0.05 wt.%. The inclination of the beam polarization plane was controlled with the use of half-wave plate **λ/2**. First, the crystal was irradiated in such a manner that the polarization vector was perpendicular to the main axis *c* of the crystal. In this case, an ordinary wave propagated in the medium. The scattering field was linearly polarized, and the polarization azimuth coincided with the azimuth of the initial beam. In 35 min, the process of scattering became stationary. Then, half-wave plate **λ/2** was turned in such a manner that the angle between the polarization vector of the laser beam and the axis c became equal to 59° (b), so that the difference between diffraction efficiencies for the ordinary and extraordinary waves became compensated. As the result, laser induced scattering speckle field becomes elliptically polarized. In such a geometry of recording and reading the noise gratings, the scattering field was elliptically polarized and had no prevailing component. The component, which corresponded to the extraordinary wave, started to record noise gratings, which competed with those recorded earlier. The process of scattering ceased to be stationary, and the speckle field began to smear. Both the intensity distribution and its polarization structure changed completely and stochastically. The scattered radiation was collimated making use of lens **L** and was analyzed with the help of Stokes-analyzer (quarter-wave plate λ/4 and the polarizer P) and a CCD- chamber. The Stokes-analyzer consisted of a quarter- wave plate (the error was about λ/100) and a polarizer with a polarization ability of 1/500. The relative error of the determination of the intensity of Stokes-parameters was induced by the imperfection of optical elements and was equal to 4%. Scattering at an angle of 6° (the analyzed section was $1.5 \times 10^{-4}$ sr.) was analyzed. To minimize unwanted changes of investigated field during measurement of all stokes intensity components laser beam was blocked by electro-mechanical shutter S on the time of plate λ/4 and polarizer P readjustments. Optical damage relaxes during many hours. Therefore, measured field topology in some cadre can be considered practically unchanged during these operations and real interval between cadres was much less then 15s and equals few seconds only.

## 3. EXPERIMENTAL RESULTS AND DISCUSSION

First measurements have show that there are two types of chain reaction trajectories: (*i*) *loop* trajectories [8] (Fig.5*a*) when nucleated hyperbolics pair annihilates soon and (*ii*) *chain* trajectories (Fig.5*b*) when born hyperbolics repel and after some intermediate tranformations to elliptics and back annihilate with *alien* partners for other chains. Hundreds of such trajectories were measured. And it was found that all of them belong to these two classes. Next task was to connect these quit different scenarios with local dynamics of developing speckle field including morphological transformations of involved C points and diabolos.

During short-term loop trajectory nucleated pair of singularities annihilates shortly (Fig.5*a*). Quite another is general scenario of each chain from total chain trajectory (Fig.5*a*). Two nucleated singularities repel and after movement annihilate separately with *alien* singularities nucleated in other space-time points. Two remained singularities from neighbor chains attract to new pair of singularities etc. Total chain trajectory reaction is limited by dimensions of illuminated volume in photorefractive crystal and kinetics of 'optical damage' development.

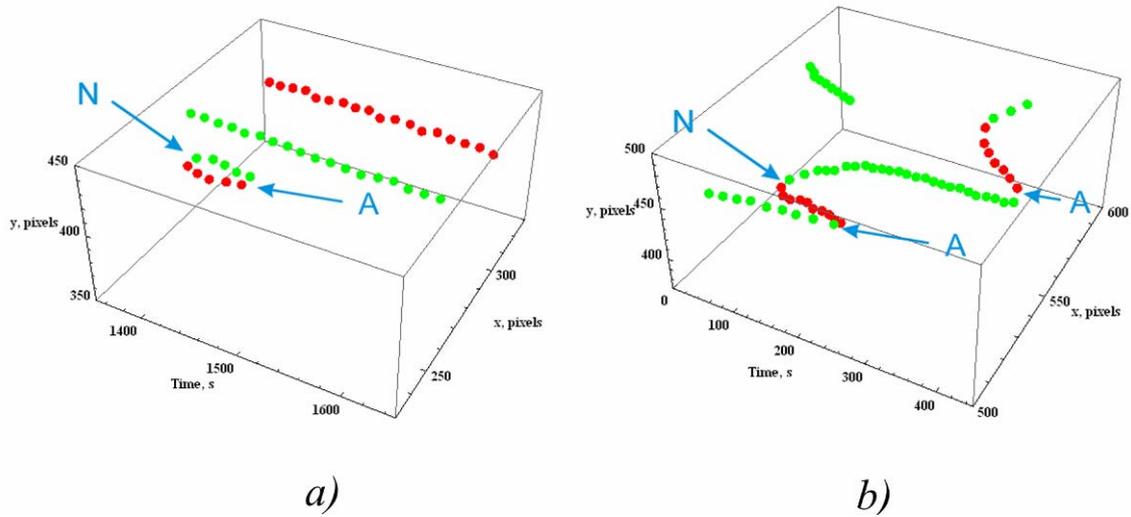

Figure 5. Two type of C points dynamics (a) loop trajectories (b) chain trajectories in 3D space-time presentation. N (A) mark moments of singularities pair nucleation (annihilation).

Detailed topological (morphological) scenarios of both types of C points trajectories is considered below through sequence of 2D contour maps with contours of equal height (20 intervals from max to min) and C points with marked morphological types of C points and types of diabolos. Loop trajectory starts on a singularities-free smooth speckle at 1350s (Fig.6*a*).

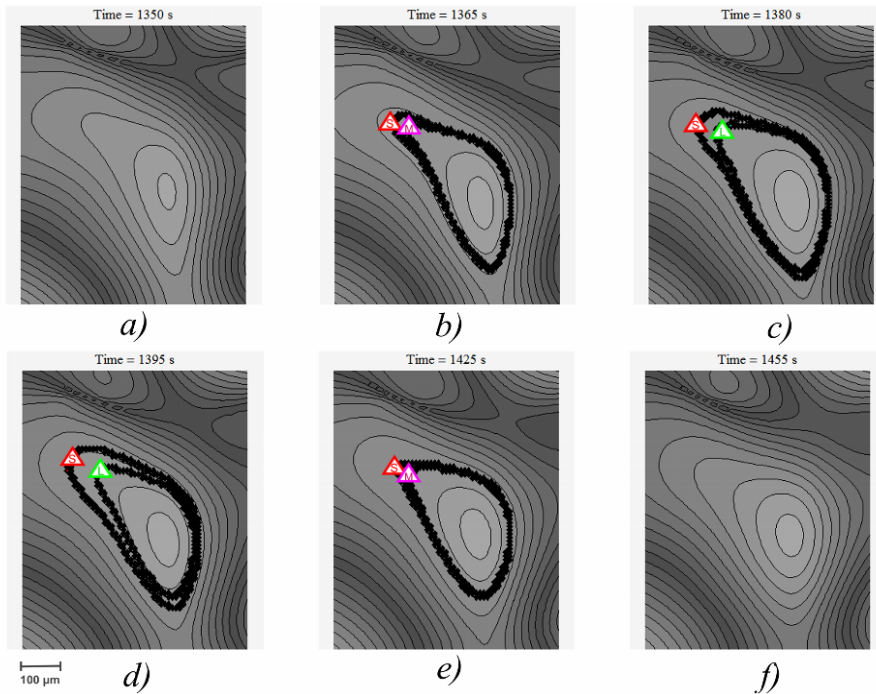

Figure 6. Topological and morphological dynamics of a loop trajectory. Time marked for each episode is counted from the start of photorefractive crystal illumination by laser beam. Triangle mark hyperbolics and letters S (star), M (monstar), L (lemon) inside it the morphological forms of C points. Thick black lines are height of contours going through C point.

Their correspondent thick level contours coincide practically far from hyperbolics. It witnesses that moment of cadre measurement coincides occasionally with hyperbolics nucleation. Born C points repel then a little, level contours are separated and H (M) transforms to H (L) according with theory predictions [2, 6] (Fig.6*c, d*). This pair attracts again shortly and lemon transforms then back to monstar (Fig.6*e*) and they annihilate finally (Fig.6*f*). The star-monstar hyperbolics pair is born on its slope at 1365s (Fig.6*b*).

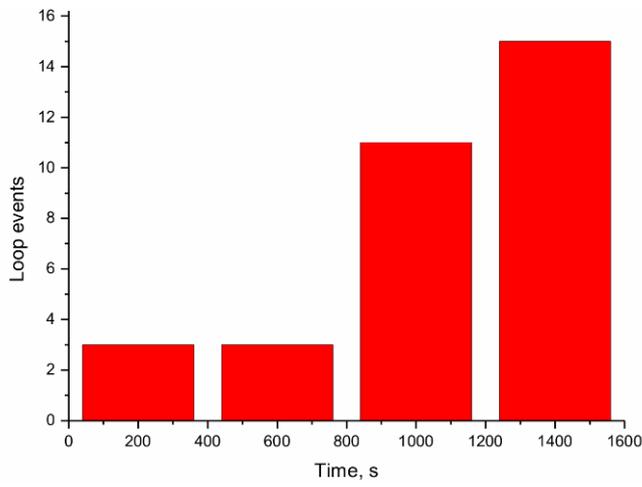

Figure 7. Time histogram of loop trajectories number.

All these transformations are happen on same slope of the speckle which remains practically unchanged during loop development. Full duration of this short life is nearly 45s only and elliptics aren't involved into loop trajectories evolution. Such scenario was valid for all loop trajectories which are rare enough. It is seen that loop trajectories didn't take part in general development of induced singularities evolution. From this point of view, they can be considered as the peculiar "topological impurities". Relative part of loop trajectories equals ~10 % of total trajectories only. But their quantity grows appreciably to the end of optical damage. Only few of them are realized on the first half of field evolution contrary to tens during period of optical damage saturation (Fig.7). Physical reasons of loops realization will be discussed later.

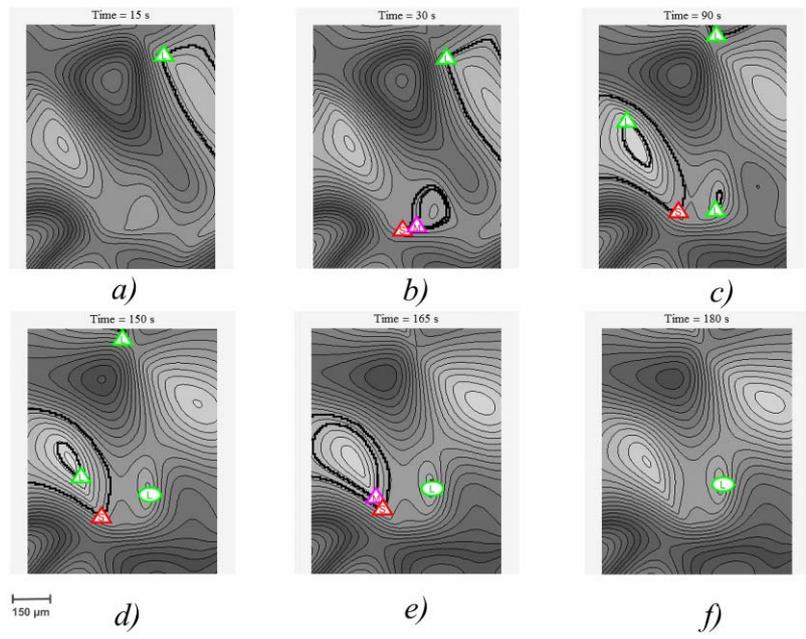

Figure 8. Dynamics of a chain topological reaction (first stage). Hyperbolic – elliptic transformations are seen.

Quite another and much more reach is topological and morphological evolution of chain reactions. Their duration is nearly few times longer. Therefore its first and second topological stages are presented in Fig.8 and Fig.9. Shown chain reaction starts on a speckle with complicated shape containing main and lower tops (Fig.8*a*). The H (L) in upper right corner belongs to other chain reaction and will not take part in considered chain reaction which started on the slope of the local low maximum in standard way by nucleation of H (S) - H (M) pair (Fig.8*b*). They repel then, H (M) transforms to H (L) in standard way and climbs to the top of lower maximum (Fig.8*c*). Simultaneously, H (M) moves to the lower right maximum and embraces it by his height contour. The upper H (L) from other chain reaction mows this time to the remained H (S) hyperbolic and right H (L) transforms to E (L) (Fig.8*d*). H (L) is moving down to H (S) and transforms to H (M) (Fig.8*e*). They annihilate finally and E (L) remains on the lower maximum of the speckle (Fig.8*f*). This finishes first part of considered complicated chain reaction when only one partner E (L) remains at 180s.

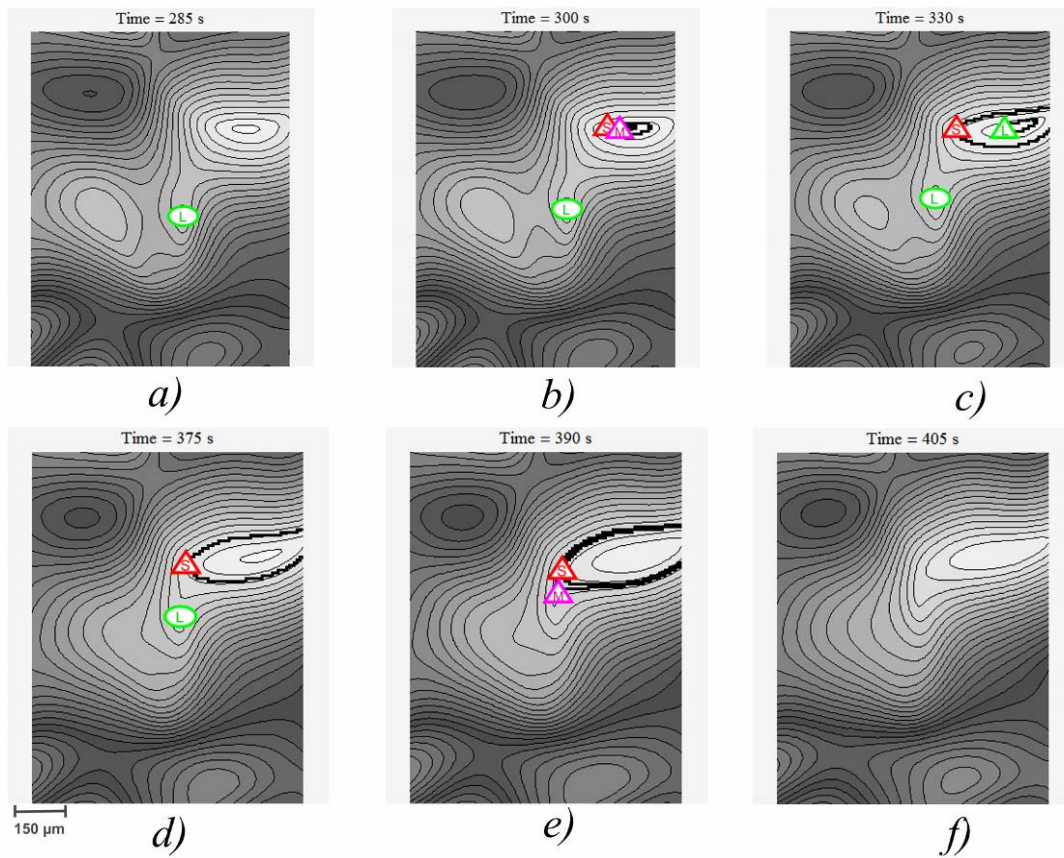

Figure 9. Dynamics of chain topological reaction (second stage)

It remains up to 285s (Fig.9*a*), when second stage of considered chain reaction started at 300s when new H (S) – H (M) pair nucleates near top of the upper right higher speckle hill above the remained E (L) diabolo (Fig.9*b*). Singularity H (M) transforms to (H (L) leaves the field of view (Fig.9*c*). The remained pair of alien E (L) and H (S) singularities E (L) repels E (L) transforms to H (M) needed for this pair of singularities annihilation (Fig.9*e*). Finally they annihilate (Fig.9*f*) finishing this peace of chain reactions. As we see, full duration of considered chain reaction equals nearly 400s, i.e. few times longer then the duration of measured loop trajectory. This development of bounded chain reactions prolongs all of time of optical damage.

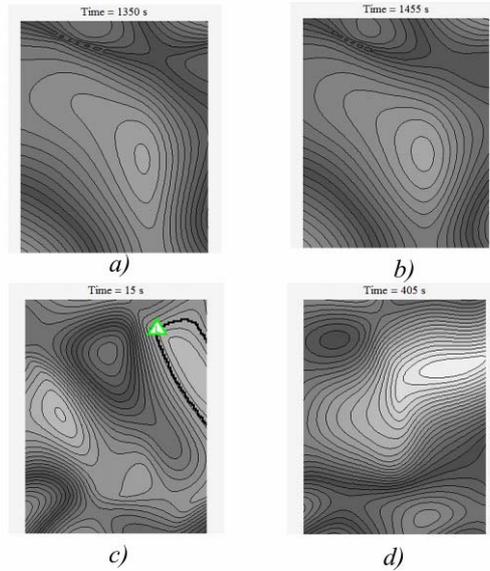

Figure 10. Comparison of initial and final configurations of the speckle field for measured loop (*a, b*) and chain (*c, d*) trajectories.

As we see, full duration of considered chain reaction equals nearly 400s, i.e. few times longer then the duration of measured loop trajectory. This development of bounded chain reactions prolongs all of time of optical damage.

Comparison of initial and final configurations of the speckle field for measured loop and chain trajectories (Fig.10) show their essential difference. Indeed, structures of speckle before and after a loop trajectory coincide practically (Fig.10a). Contrary, structure of speckles preceding the considered chain reaction and final one possess quit different structure (Fig.10b). This reflects strong perturbation of speckle field topological structure during development of optical damage in LiNbO3.

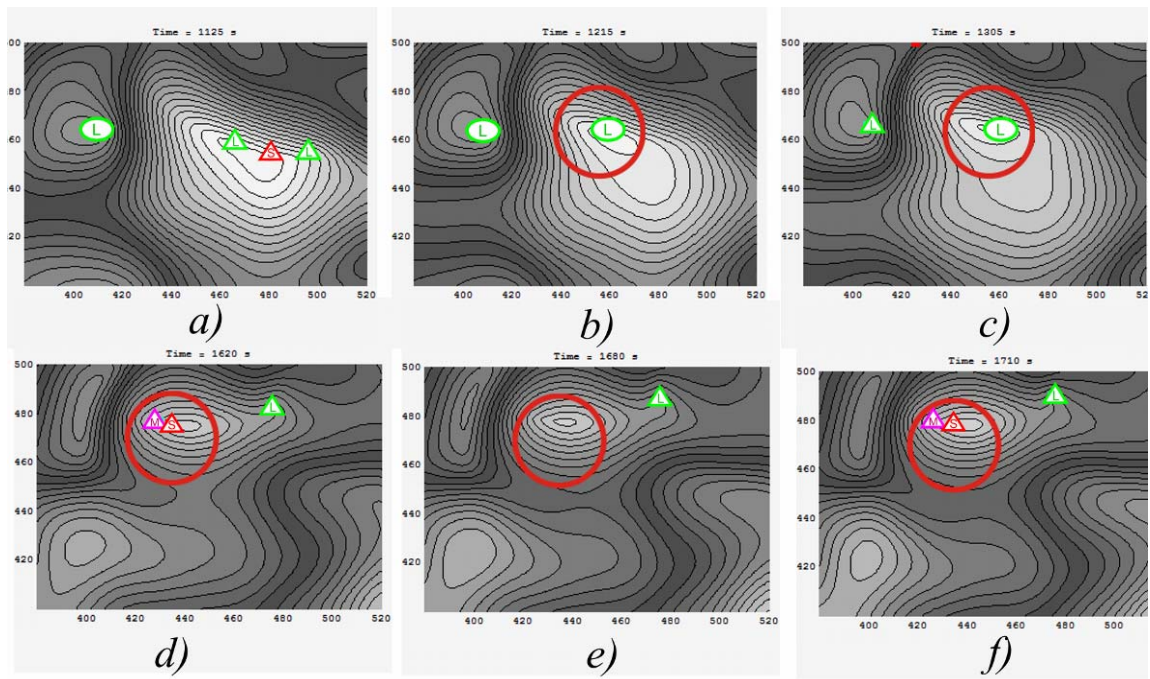

Figure 11. 'Development of the 'wave-like' loop trajectory in one 'stability zone' around speckle maximum marked by circles.

At the same time, there are '*stability zones*' during speckle field dynamics for loop reaction (Fig.11). Here is shown rear event development of the 'wave-like' loop trajectory in one 'stability zone' around speckle maximum.

As it is seen, four C points exist at 1125s (Fig.11*a*). After 90s (Fig11*b*) only two E (L) elliptics remain located on the neighbor local maxima. This topological structure didn't change practically during next 90s (Fig11*c*). And after more then 300s pair H (S) and H (M) appeared (Fig11*d*) which annihilate before 1680s (Fig11*e*). This pair reappeared at 1710s (Fig11*f*). As we see mark zone stability holds nearly 500s, the very long time for topological reactions which change their topology during tens of seconds. Another example of chain reaction with zone of stability is shown in Fig.12.

It is interesting that this behavior is typical for all loop trajectories during development of optical damage in photorefractive crystal. This witnesses that there are few small volumes inside photorefractive crystal with refractive indices smooth distribution along with very inhomogeneous distribution of laser field inside main part of the crystal due to heterogeneities of its refractive indices. Namely these parts of crystal produce zones of stability inside loop trajectories. Grows of loop trajectories quantity to the end of optical damage processes (Fig.6) is natural due to saturation of these processes. It's amazing that such stability zones exist both in loop and long chain reaction (Figs.11, 12). They exist on the tops or their near vicinities and are of the same origin.

Few words about realized diabolos and statistics of diabolos forms (H, E) and C points morphologies (S, M, L) in chain and loops trajectories. The generic nucleation (annihilation) of hyperbolics pairs HH was observed only. Some upper hyperbolics can climb on the nearest tops and transformed to elliptics. All nucleated C points pairs are S and M combinations only [6]. When born C pair repel, monstar M transforms to lemon L. From this follows that quantity of nucleated stars $N_S$ equal sum of $N_M$ and $N_L$. Experiments with chain reactions confirm this prediction (see also [6]). It is interesting that all singularities born during optical damage exist near to top of same speckles. It follows that topologically allowed H' diabolos existing around minima of wave front [13] were not realized in investigated class of topological reactions.

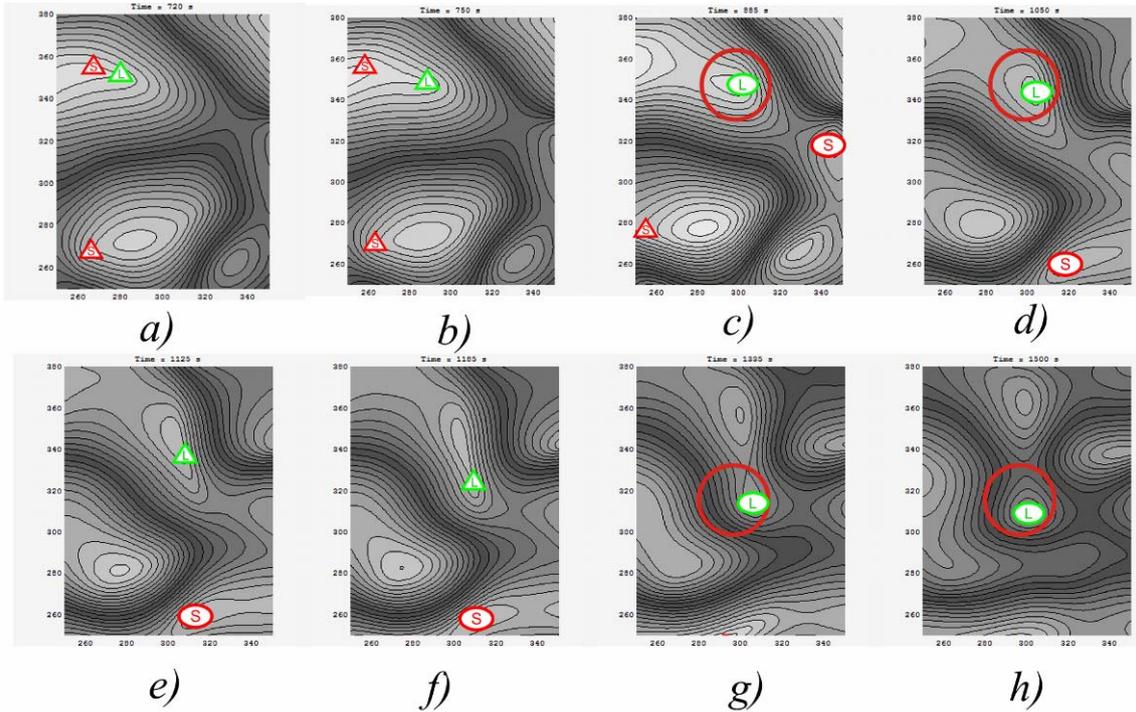

Figure 12. Stability zone during development of a long-term chain reaction marked by circles.

One of points of singular optics of complicated is the statistics of diabolos and C point forms. It was considered firstly for isotropic random fields by M. Dennis [16] and found that fraction of H (S), H (L), E (S) and E (L) is nearly equal and amount main part of all singularities. The fractional densities of three morphological forms of C points equal α(S) = 0.5, α(M) = 0.05279, and α(L) = 0.44721. Real fractions of hyperbolic and elliptic diabolos in developing speckle fields is quite another and dictated by lows of singularities pair nucleation, transformation, and annihilation (Table 1). As we have observed, all events of singularities nucleation and annihilation proceed generically as H (S) – H (M) pairs. Elliptics can be realized through H → E transformation only. Therefore, they are absent in loop trajectories and are seldom enough during chain reaction. Much above is the fraction of monstars which appear at the moment of all nucleation and annihilation events of hyperbolics pairs.

Table 1. Statistical weight of C points

| C point type | Theory [16] | Ful life-cycle (experiment) | Nucleation-Annihilation (experiment) |
|---|---|---|---|
| E(S) | 0.25 | 0.23 ± 0.015 | 0 + 0.015 |
| H(S) | 0.25 | 0.27 ± 0.015 | 0.5 ± 0.015 |
| E(L) | 0.2347 | 0.24 ± 0.015 | 0 + 0.015 |
| H(L) | 0.2125 | 0.20 ± 0.015 | 0 + 0.015 |
| E(M) | 0.0152 | 0.018 ± 0.015 | 0 + 0.015 |
| H(M) | 0.0376 | 0.039 ± 0.015 | 0.5 ± 0.015 |

Finally, let we discus firstly the effect of instantaneous topological rearrangement of light field topology during nucleation (annihilation) of polarization singularities pair and their transformations. The real topological structure of any ordinary speckle without singularities each moment of speckle field generic development is the result of mainly constructive interference of field components with random amplitude, polarization and amplitude. Therefore, azimuth of resulting polarizations ellipses, their shape and values of amplitudes possess no special elements of symmetry. But when singularities pair is nucleated as H (S) – H (M) pair [2, 6], morphological and topological structure of this speckle changes instantly (more precisely during period of light vibrations): ellipses major axes are oriented along three lines according S and M morphological forms, values of major and minor axes form two hyperbolic surfaces with common center in C point [9]. This up to now not marked effect can be classified as topological (morphological) instant self-action of developing generic speckle field. We plan measure it in near future be comparison of fixed parameters distribution in the speckle "pregnant" with singularities and its form parameters after singularities nucleation. The same origin more fine transformations have to be realized during change of polarization topology and morphology.

## 4. CONCLUSION

Main physical findings can be summarized as follows:

- The generic naturally developing speckle field created and driven by 'optical damage' effect in photorefractive crystals under laser illumination is single up to moment unique polygon for fulfill regularities of dynamic singular optics.

- Evolution of speckle field proceeds by two types of trajectories: (*i*) more rare (~ 10%) *loop* trajectories (*ii*) and most probable (~ 90%) *chain* trajectories. All of them start and finish by nucleation and annihilation of hyperbolics pair H (S) – H(M). All aloud combinations of diabolos topological forms (hyperbolic, elliptic) and morphological form of C points (star, monstar, lemon) are realized.

- The short-time loop trajectories proceed by repel a little of nucleated hyperbolics, their back attraction and annihilation. Loop trajectories don't participate in general evolution of speckle field and can be treated as 'topological defects' in stream of chain reactions.

- The long-time chain trajectories proceed by essential moving off of nucleated hyperbolics pair, transformation of upper hyperbolics to elliptic, few morphological transformations of C point, and annihilation *alien* singularities from next chain reactions. Due to relay race of such nucleation and annihilation events totality of such chain reaction total chain reaction lasted all time of 'optical damage' effect.

- Both types of reactions contain 'islands of stability' due to local saturation of optical damage process. Their density grows to the end of optical damage development.

- In general, topological (morphological) dynamics obey each moment the principle of minimal potential energy for open system.

It is natural to conclude that established regularities have to valid for any variant of singular light field nonlinear development.

## ACKNOWLEDGMENT

Authors are thankful to Prof. I. Freund for helpful discussions.